# High-resolution microwave frequency dissemination on an 86-km urban optical link

**O. Lopez[1], A. Amy-Klein[1], M. Lours[2], Ch. Chardonnet[1] and G. Santarelli[2]**

[1] Laboratoire de Physique des Lasers, Université Paris 13, CNRS, 99 Av. J.B. Clément, 93430 Villetaneuse, France

[2] LNE-SYRTE, Observatoire de Paris, CNRS, UPMC, 61 Av. de l'Observatoire, Paris, France

Abstract : we report the first demonstration of a long-distance ultra stable frequency dissemination in the microwave range. A 9.15 GHz signal is transferred through a 86-km urban optical link with a fractional frequency instability of $1.3 \times 10^{-15}$ at 1 s integration time and below $10^{-18}$ at one day. The optical link phase noise compensation is performed with a round-trip method. To achieve such a result we implement light polarisation scrambling and dispersion compensation. This link outperforms all the previous radiofrequency links and compares well with recently demonstrated full optical links.

corresponding author Giorgio Santarelli, email : giorgio.santarelli@obspm.fr

# 1. Introduction

Ultra-stable frequency transfer between remotely located laboratories is required in time and frequency metrology, fundamental physics, particle accelerators and astronomy. Distant clock comparisons are currently performed using satellites, by Two-Way Satellite Time and Frequency Transfer, or through the Global Positioning System. Both methods are limited to a $10^{-15}$ fractional frequency instability for one day of averaging time [1]. This is insufficient to transfer the properties of modern cold atom microwave frequency standards which have demonstrated frequency instability of a few $10^{-16}$ at one day [2, 3]. Beyond metrology, high-resolution clock comparison is essential for advanced tests in fundamental physics, such as tests of the stability of fundamental constants [4-6].

To overcome current free space link limitations, the transmission of standard frequencies over optical fibers has been investigated for several years [7-10]. This technique takes advantage of the low attenuation, high reliability and continuous availability of fibers.

Radio frequency (RF) transmission using amplitude modulation of an optical carrier at 1 GHz have demonstrated a frequency instability as low as $5\times10^{-15}$ at 1 s and $2\times10^{-18}$ at one day over 86 km [11]. Direct optical frequency transfer [12-15] can provide even better stability and be extended to greater distances. For both methods a phase noise correction is needed to compensate for the fluctuation of the propagation delay due to mechanical perturbation and temperature variation along the fibre. For this purpose the so-called round-trip method is used.

Radio frequency transmission over an optical link has already been demonstrated at 100 MHz, 1 GHz and a few tens and hundreds of GHz [8, 9, 11, 16, 17]. In this paper, we report on the transmission of a microwave frequency reference signal at 9.15 GHz over a 86-km urban fiber link connecting our two laboratories, LPL and LNE-SYRTE. This work pursues the development of stable frequency distribution of a reference signal already demonstrated between our two laboratories in the RF domain at 100 MHz [8] and 1 GHz [11]. The fiber optical length fluctuations induce phase fluctuations proportional to the modulation frequency, thus moving to higher frequency potentially leads to an increase of the signal-to-noise ratio of the detected fiber phase fluctuations. Moreover a microwave frequency of about 10 GHz is well suited to applications concerning particle accelerators [18] and astronomy and is close to the 9.192 GHz caesium transition frequency used for the definition of the SI second. In the following, we describe the new set-up. Then we present the resulting performance, discuss the limitations and conclude.

## 2. Experimental set-up

The schematic of the optical link and compensation system is displayed on Figure 1. The link is obtained by cascading two 43-km twin fibers connecting the laboratories LPL and LNE-SYRTE. Both ends are collocated at LPL, one labelled as local and the other as remote.

The performance of our previous optical link was limited by the signal-to-noise ratio at detection and the phase noise and long-term instability of the laser source which induced a parasitic noise due to the fiber dispersion [11]. Thus two main changes have been implemented. The reference carrier frequency has been increased from 1 GHz to 9.15 GHz inducing a complete change of the compensation system, laser sources and detection set-up. Moreover a section of negative dispersion fiber has been added to the link to compensate for its dispersion.

At each end two very low phase noise microwave synthesis systems generate all the signals necessary for the compensation system. At the local end, a 9.15 GHz Yttrium Iron Garnet microwave oscillator (YIG) is phase-locked to a low phase noise 100 MHz voltage controlled quartz oscillator (VCXO) with a double stage RF down conversion approach using a 200 MHz sampling mixer [19, 20]. At the remote end a similar technique is used to phase lock a second microwave oscillator operating at 9.25 GHz by harmonic sampling to a low phase noise 1 GHz Surface Acoustic Wave oscillator (SAW). Each microwave signal directly modulates the beam intensity of a 6 mW distributed feedback laser diode (DFB-LD) at 1.55 µm using an Electro-Absorption Modulator (EAM), the DFB-LD and the EAM being integrated on the same chip (Mitsubishi FU-653SEA). The modulation power of about 5 dBm is injected in the EAM using a bias-tee to optimize the EAM operation point with a DC bias of a few Volts. The modulation is detected with a fast photodiode (10 GHz bandwidth Discovery Semiconductors DSC50S) at the other end of the fiber. We measured the phase noise of the emitter/receiver part (DFB-LD, photodiode and microwave amplifier) to be $-105$ dBrad$^2$/Hz at 1 Hz offset, rolling down with a close to $1/f$ slope up to 1 kHz. This residual noise is compatible with a frequency instability better than $10^{-15}$ at 1 s. At the remote end the 9.25 GHz microwave signal is mixed with the incoming 9.15 GHz signal detected by the photodiode and the resulting beat-note signal at 100 MHz is mixed to DC with the SAW oscillator output digitally divided by ten. This error signal is used to phase lock the SAW oscillator to the incoming signal and consequently the 9.25 GHz with a bandwidth of a few kHz. This way the 9.25 GHz YIG oscillator signal reproduces the one-way phase perturbations of the link. This signal modulates a second laser diode which is injected back into the link using an optical circulator. At the local end the 9.25 GHz backward signal is detected with a second fast photodiode (DSC50S). This signal carries the round-trip phase perturbations accumulated over the link. In order to compensate for these perturbations the phase is revealed by successive mixing processes with the input 9.15 GHz signal and the 100 MHz from the VCXO. This generates an error signal which is processed by a simple low frequency loop filter and applied to two variable delay lines at the local input in order to cancel out the link phase perturbations.

One variable delay line is home made with a 15-meter long optical fiber wrapped around a 20-mm diameter cylindrical piezoelectric transducer (PZT) to correct the fast and small phase perturbations [17]. Voltage applied to the PZT stretches the fiber with a dynamic range of 15 ps and a bandwidth of about 1 kHz. The second variable delay line consists of a 2.5-km long fiber spool in an enclosure which is temperature controlled by thermoelectric modules. This device corrects slow and large phase perturbations with a dynamic range of about 4 ns corresponding to a temperature range of 50°C (88 ps/°C). It enables the correction of a global temperature change of 1.5°C of the optical link, whereas the temperature change is typically less than 0.5 °C per week.

The use of two different microwave frequencies for the forward and backward signals (9.15 GHz and 9.25 GHz) prevents interferences between the main signal, Brillouin backscattering and parasitic reflections from connectors and splices along the link.

The polarization mode dispersion (PMD) and the chromatic dispersion are two detrimental fiber propagation effects that limit the performance of the correction system as discussed in [11, 21]. Both phenomena induce an asymmetry of the phase perturbation in forward and backward propagation directions along the link with the result that the round-trip phase fluctuations do not exactly correspond to twice the forward phase fluctuations. To minimize PMD, the laser beam polarization is directly scrambled at each emitted source by a 3 axes polarization scrambler at frequencies higher than the inverse of the light propagation delay in the overall round-trip (0.5 ms). Each scrambler acts as three cascaded variable retardation waveplates excited at different resonant frequencies (approximately 60 kHz, 100 kHz and 130 kHz). This enables the exploration of all polarization states. The chromatic dispersion of the fiber ($D \sim -17$ ps/km/nm) converts the laser diodes' frequency noise into an excess phase noise of the optical link [11]. To reduce this effect 11 km of highly negative dispersion fiber is inserted at the local end. Thus the total dispersion is reduced to less than 5% of the original 86-km fiber dispersion. This also prevents periodic signal fading and extinction along the link. Without this compensation the dispersion creates a differential phase shift between the microwave sidebands of the optical signal. It amounts to $\Delta\phi = 2\pi c D L (\Omega/\omega)^2$ with $D$ the fiber dispersion, $L$ the fiber length, and $\Omega$ and $\omega$ the microwave and optical angular frequencies respectively. This leads to a modulation of the detected microwave signal amplitude with the fiber length with a first complete extinction at around 22 km [11].

Two optical amplifiers (EDFA) are used to optimize the signal-to-noise ratio at detection and to compensate for the 11 dB additional optical losses introduced by the negative dispersion fiber. Despite these EDFAs and additional attenuation, the signal-to-noise ratio has not been degraded thanks to the higher modulation frequency compared to a 1 GHz system [11].

## 3. Results and discussion

We measure the fractional stability of the compensated link by analyzing the phase variation between the local end 9.15 GHz and the remote end 9.25 GHz (Fig. 1). The beat signal at 9.25 GHz is down-converted to DC by mixing with the 9.15 GHz input and the 100 MHz LO. Figure 2 shows the residual phase noise spectral density of the compensated 86-km optical link (red trace) measured at 9.25 GHz, the system noise floor obtained after replacing the fiber link with an equivalent optical attenuator (black dotted trace) and the noise of the laser diode, photodiode and amplifier (blue dashed trace). The latter noise does not limit the compensation system. The red trace shows a rather dense distribution of spurious bright lines due to the polarization scrambling process and their multi frequency mixing products converted into phase variations. These unwanted lines are not a limiting factor for distant clocks comparisons. However, in order to transfer an ultra-stable oscillator, it is necessary to remove the spurious lines beyond 50-100Hz. This could be done by locking a commercially available low phase-noise oscillator to the transmitted signal with a bandwidth around 50 Hz and reducing its phase-noise close to the carrier.

Figure 3 shows the temporal behaviour of the link propagation delay in open and closed loop over 4 days. The open loop signal was obtained simultaneously with the closed loop phase by measuring the temperature of the 2.5 km heated spool used for the correction and converting it in phase. The free-running fiber propagation delay spans over about 2 ns while the closed loop signal is confined well below 200 fs peak-to-peak. This demonstrates that the correction systems have a rejection factor close to $10^4$. Figure 4 shows the Allan deviation calculated from the propagation delay data measured on the 86-km compensated link. A 15 Hz low-pass filter is used to remove the spurious lines and the high frequency phase noise of the microwave signal before phase sampling. We obtain a frequency instability of the link of $1.3 \times 10^{-15}$ at 1 s integration time and much better than $10^{-18}$ at one day integration time (red circles). This result is compared with previous results at 1 GHz (blue squares) [11] together with the free running frequency stability of the link (black diamonds). This demonstrates that we have significantly improved the frequency stability of the link by using amplitude modulation at higher microwave frequencies and minimizing the unwanted fiber propagation effects. A linear fit calculated over the entire dataset show a frequency bias of about $2 \times 10^{-19}$, compatible with zero within the errors bars.

The ultimate link stability is still an open question and we need to analyze the different sources limiting the stability. To demonstrate the PMD deleterious effects on the fiber propagation, we have removed the polarization scrambler at the remote end and performed a phase measurement.

In this case the long-term stability of the link is severely degraded (black squares in figure 5) and reaches a Flicker floor at around $10^{-17}$. We have also replaced the negative dispersion fiber with an equivalent optical attenuator. In this case both short-term and long-term stability are affected (blue triangles in figure 5). The compensation system noise floor is shown in figure 5 (grey diamonds) where the 86-km link has been replaced by an equivalent optical attenuator. Below 2000 s, there is a good agreement between the 86-km stability link (red circles) and the compensation system floor. The signal-to-noise ratio at the optical detection is limiting this noise floor at short term. The frequency instability scales with approximately a $\tau^{-2/3}$ slope. This noise level has been obtained with a careful control of the thermal environment of the electronics package and the uncommon sections of the optical paths. Over the long term, the optical link stability is no longer limited by the compensator floor. This long-term instability can be attributed to residual PMD effects or amplitude modulation to phase modulation (AM/PM) conversion in photodiodes and mixers [22, 23].

## 4. Conclusion.

We have demonstrated an 86-km compensated optical link using an urban telecom network with a frequency instability of $1.3 \times 10^{-15}$ at 1 s and below $10^{-18}$ after one day of integration time. These results were obtained due to the use of two slightly different modulation frequencies for the two different propagation directions, the scrambling of the light polarization state and the compensation of the fiber chromatic dispersion. The ultimate limitation of the compensator itself is reached and can not be overcome easily. The stability of the present system enables comparison of the best frequency standards, both in the microwave and optical region, over distances of up to 100 km. For longer distances, attenuation along the fiber is a crucial limitation. Optical amplifiers, which limit the frequency instability at a level slightly below $10^{-14}$ at 1 GHz, have a negligible effect at 10 GHz and can be used to recover a sufficient signal to noise ratio for distances up to about 200 km. For longer spans, an all optical compensation system is more promising. Modern dispersion shifted fibers with lower PMD values allow even better performance and can potentially extend the link beyond 200 km.

**Acknowledgments**

We acknowledge funding support from the Ministère de la Recherche and European Space Agency/ESOC.

**References**

[1] A. Bauch, J. Achkar, S. Bize, D. Calonico, R. Dach, R. Hlavac, L. Lorini, T. Parker, G. Petit, D. Piester, K. Szymaniec, P. Uhrich, "Comparison between frequency standards in Europe and the USA at the $10^{-15}$ uncertainty level", Metrologia, 43, 109-120 (2006).

[2] C. Vian, P. Rosenbusch, H. Marion, S. Bize, L. Cacciapuoti, S. Zhang, M. Abgrall, D. Chambon, I. Maksimovic, P. Laurent, G. Santarelli, A. Clairon, A. Luiten, M. Tobar and C. Salomon, "BNM-SYRTE fountains: recent results" *IEEE Transactions on Instrumentation and Measurement*, 54, 833-836 (2005).

[3] S. Weyers, B. Lipphardt, and H. Schnatz, "Reaching the quantum limit in a fountain clock using a microwave oscillator phase locked to an ultrastable laser", Phys. Rev. A, 79, 031803R (2009).

[4] J. P. Uzan, "The fundamental constants and their variation: observational and theoretical status", Rev. Mod. Phys. 75, 403-455 (2003).

[5] S. G. Karshenboim, "Fundamental physical constants: looking from different angles", Can. J. Phys., 83, 767 (2005).

[6] V. V. Flambaum, "Enhanced effect of temporal variation of the fine-structure constant in diatomic molecules", Phys. Rev. D, 69, 115006 (2004).

[7] A. Amy-Klein, A. Goncharov, C. Daussy, C. Grain, O. Lopez, G. Santarelli, C Chardonnet, "Absolute frequency measurement in the 28-THz spectral region with a femtosecond laser comb and a long-distance optical link to a primary standard", Appl. Phys. B 78, 25-30 (2004).

[8] F. Narbonneau, M. Lours, S. Bize, A. Clairon, G. Santarelli, O. Lopez, C. Daussy, A. Amy-Klein, C. Chardonnet, "High resolution frequency standard dissemination via optical fiber metropolitan network", Rev. Sci. Instrum., 77, 064701 (2006).

[9] M. Calhoun, S. Huang, and R. L. Tjoelker, "Stable Photonic Links for Frequency and Time Transfer in the Deep-Space Network and Antenna Arrays", Proc. of the IEEE, Special Issue on Technical Advances in Deep Space Communications & Tracking, 95, 1931-1946 (2007).

[10] S. M. Foreman, K. W. Holman, D. D. Hudson, D. J. Jones, and J. Ye, "Remote transfer of ultrastable frequency references via fiber networks", Rev. Sci. Instrum., 78, 021101 (2007)

[11] O. Lopez, A. Amy-Klein, C. Daussy, Ch. Chardonnet, F. Narbonneau, M. Lours, and G. Santarelli, "86-km optical link with a resolution of $2\times10^{-18}$ for RF frequency transfer", Eur. Phys. J. D, 48, 35-41 (2008).

[12] N. R. Newbury, P. A. Williams, W. C. Swann, "Coherent transfer of an optical carrier over 251 km", Opt. Lett., 32, 3056 (2007). See also P. A. Williams, W. C. Swann, and N. R. Newbury, "High-stability transfer of an optical frequency over long fiber-optic links", J. Opt. Soc. Am. B, 25, 1284 (2008).

[13] M. Musha, F. Hong, K. Nakagawa, and K. Ueda, "Coherent optical frequency transfer over 50-km physical distance using a 120-km-long installed telecom fiber network", Opt. Express 16, 16459 (2008)

[14] H. Jiang, F. Kéfélian, S. Crane, O. Lopez, M. Lours, J. Millo, D. Holleville, P. Lemonde, Ch. Chardonnet, A. Amy-Klein, G. Santarelli, "Long-distance frequency transfer over an urban fiber link using optical phase stabilization", J. Opt. Soc. Am. B, 25, 2029 (2008) and F. Kéfélian, O. Lopez, H. Jiang, Ch. Chardonnet, A. Amy-Klein and G. Santarelli, "High-resolution optical frequency dissemination on a telecommunication network with data traffic", Opt. Lett, 34, 1573-1575 (2009).

[15] G. Grosche, O. Terra, K. Predehl, R. Holzwarth, B. Lipphardt, F. Vogt, U. Sterr, and H. Schnatz, "Optical frequency transfer via 146 km fiber link with $10^{-19}$ relative accuracy", Opt. Lett., 34, 2270-2272 (2009).

[16] M. Fujieda, M. Kumagai, T. Gotoh, M. Hosokawa, "Ultrastable Frequency Dissemination via Optical Fiber at NICT", IEEE Transactions on Instrumentation and Measurement, 58, 1223-1228 (2009) and M. Kumagai, M. Fujieda, S. Nagano, and M. Hosokawa, "Stable radio frequency transfer in 114 km urban optical fiber link," Opt. Lett. **34**, 2949-2951 (2009)

[17] B. Shillue, S. AlBanna and L. D'Addario, "Transmission of low phase noise, low phase drift millimeter-wavelength references by a stabilized fiber distribution system", *Proc. IEEE Int. Top. Meet. Microw. Photon. (MWP 2004)*, 201-204, (2004).

[18] R. Wilcox, J. M. Byrd, L. Doolittle, G. Huang, and J. W. Staples, "Stable transmission of radio frequency signals on fiber links using interferometric delay sensing," Opt. Lett. **34**, 3050-3052 (2009)

[19] G. D. Rovera, G. Santarelli, and A. Clairon, "A frequency synthesis chain for the atomic fountain", IEEE Trans on Ultra. Ferro. Elec. Freq. Contr., 43, 354-358 (1996).

[20] D. Chambon, M. Lours, F. Narbonneau, M E. Tobar, A. Clairon and G. Santarelli "Design and realization of a flywheel oscillator for advanced time and frequency metrology", Rev. Sci. Instr., 76, 094704 (2005).

[21] Pengbo Shen, N. J. Gomes, W. P. Shillue, S. Albanna, "The Temporal Drift Due to Polarization Noise in a Photonic Phase Reference Distribution System", *Journal of Lightwave Technology* 26, 2754-2763 (2008)

[22] D. Eliyahu, D. Seidel, L. Maleki, "RF Amplitude and Phase-Noise Reduction of an Optical Link and an Opto-Electronic Oscillator," *IEEE Transactions on Microwave Theory and Techniques*, 56, 449-456 (2008).

[23] G. Cibiel, M. Régis, E. Tournier and O. Llopis, "AM noise impact on low level phase noise measurements," IEEE Transactions on Ultrasonics, Ferroelectrics, and frequency control 49, 784-788 (2002); see also L. M. Nelson and F.L. Walls, "Environmental effects in mixers and frequency distribution systems," Proceedings of IEEE Frequency Control Symposium, p 831-837 (1992), DOI: 10.1109/FREQ.1992.269954, (1992).

**Figure captions**

Fig. 1. Full scheme of the microwave frequency transfer; SAW: Surface Acoustic Wave oscillator; PLL: phase-locked loop; OC: optical circulator; EDFA : Erbium Doped Fiber Amplifier; EAM DFB LD : Electro Absorption Modulator Distributed Feedback Laser Diode, YIG Yttrium Iron Garnet microwave oscillator.

Fig. 2. Phase noise spectral density measured at 9.25 GHz of the compensated link (red trace), the system noise with an optical attenuator to shorten the link (black dotted trace) and the laser diode and photodiode noise (blue dashed line).

Fig. 3. Temporal behaviour of the propagation delay of the free-running optical link (smooth black trace, one data point every 100s) and the end-to-end phase of the compensated link (sampled every 100s, light grey trace).

Fig. 4. Fractional frequency stability of (a) the free running 86-km link (black diamonds), (b) 1 GHz compensation system [11] (blue squares), and (c) the compensated link at 9.15 GHz (red circles, 15 Hz measurement frequency bandwidth).

Fig.5. Fractional frequency instability of the 86-km link, in various configurations to highlight the different limitations of the system: without dispersion compensation fiber (blue triangles), forward polarization scrambling only (black squares), compensated 86-km link (red circles), system noise floor with an optical attenuation to shorten the link (grey diamonds).

Figure 1

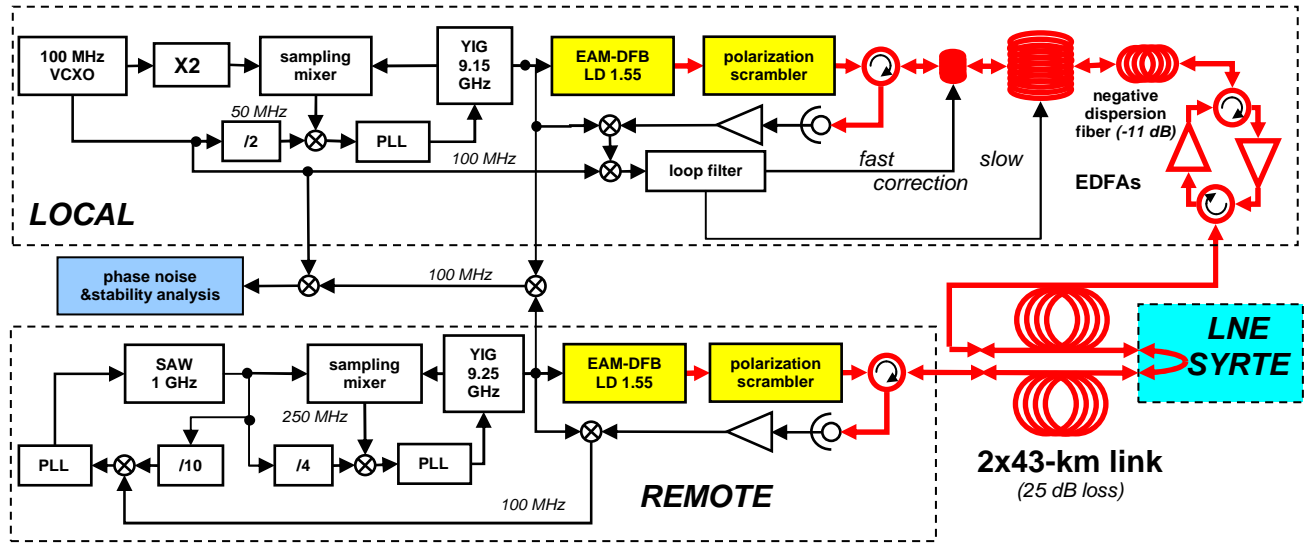

Figure 2

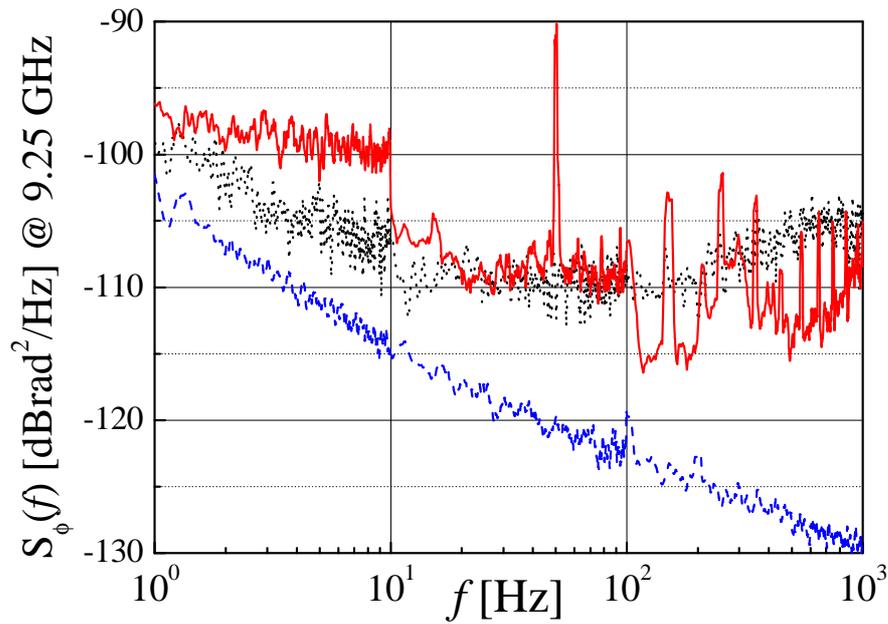

Figure 3

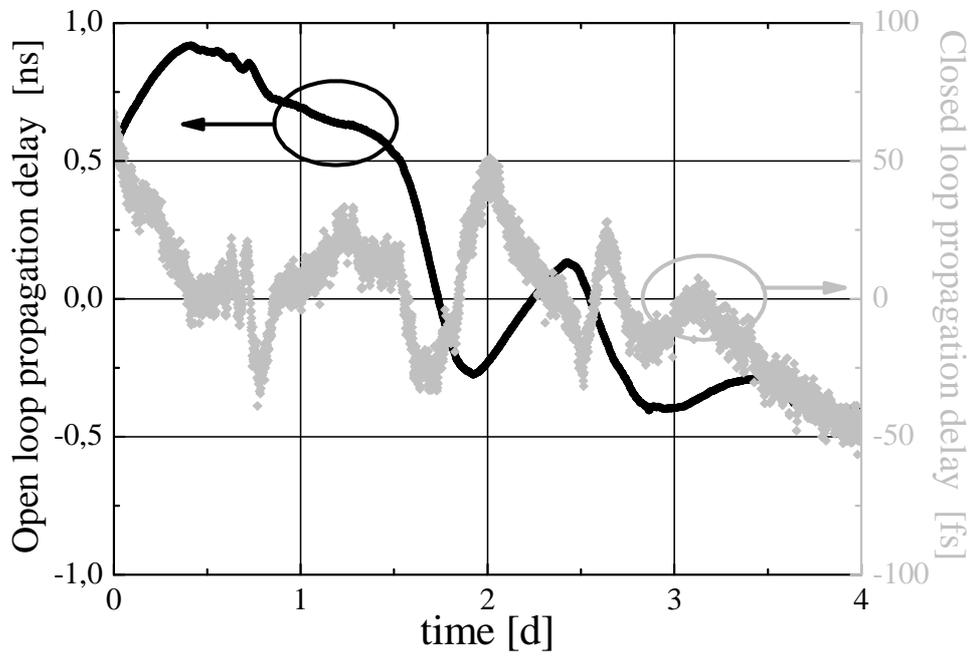

Figure 4

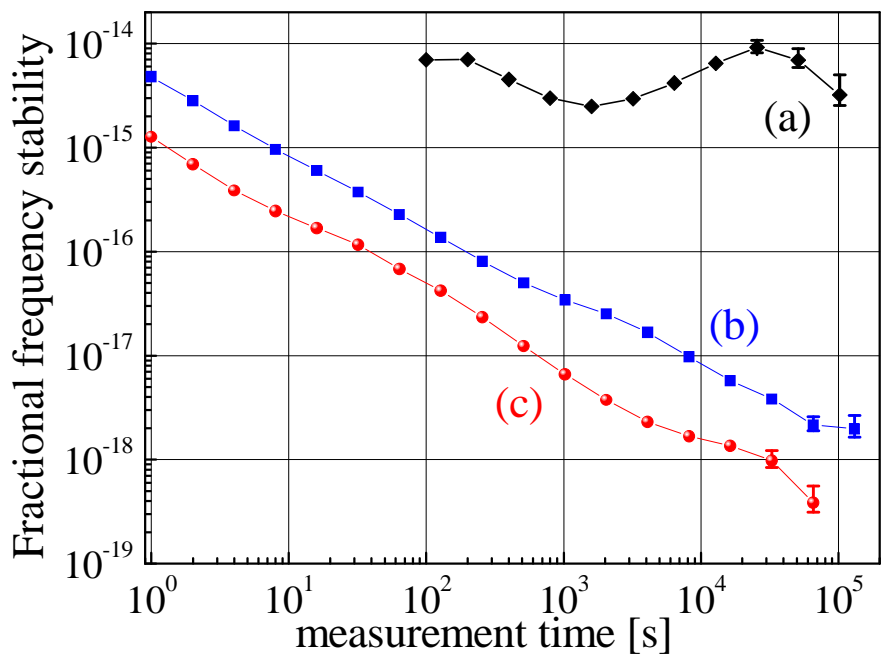

Figure 5

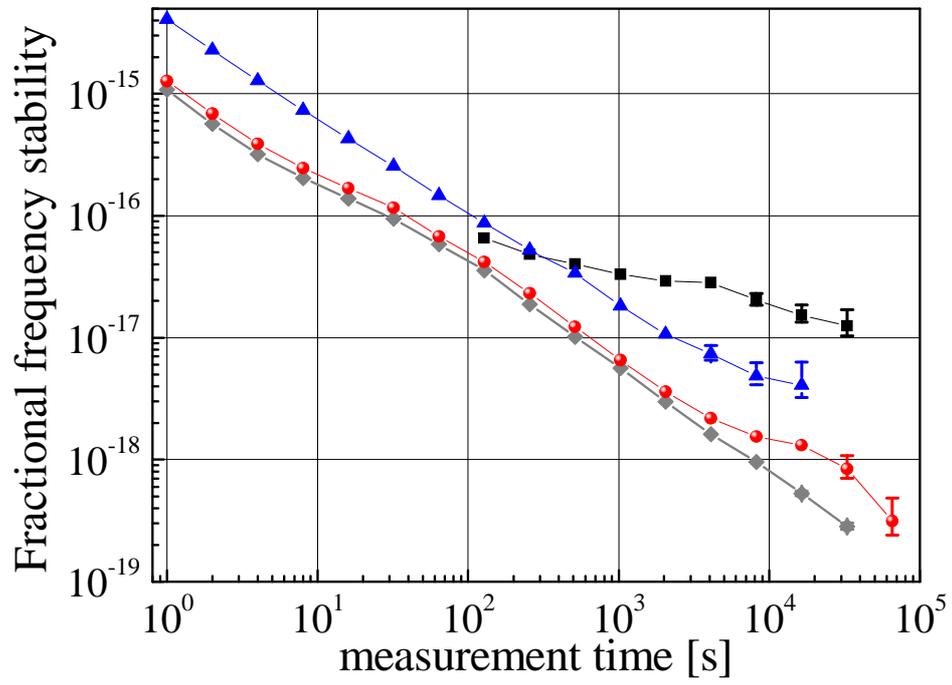